\documentclass[%
reprint,
nofootinbib,
 amsmath,amssymb,
 aps,
]{revtex4-1}

\usepackage{dcolumn}% Align table columns on decimal point
\usepackage{bm}% bold math

\usepackage[T1]{fontenc}
\usepackage[latin9]{inputenc}
\usepackage{geometry}
\usepackage{refstyle}
\usepackage{amsmath}
\usepackage{amssymb}
\usepackage{graphicx}

\usepackage[english]{babel}

\usepackage[normalem]{ulem}
\usepackage{etoolbox}
\usepackage{cancel}

\usepackage{xcolor}

\newtoggle{editing}
	\toggletrue{editing}
\iftoggle{editing}{%
	\newcommand{\BCcom}[1]{{\textbf{\textcolor{magenta}{#1 -- BC}}}}
	\newcommand{\BCout}[1]{{\textbf{\textcolor{magenta}{\sout{#1}}}}}

	\newcommand{\JSout}[1]{{\textbf{\textcolor{blue}{\sout{#1}}}}}
	\newcommand{\JSeqout}[1]{{\mathbf{\textcolor{blue}{\cancel{#1}}}}}
	\newcommand{\JSc}[1]{{\textbf{\textcolor{blue}{[#1 -- JS]}}}}

	\newcommand{\LLc}[1]{{\textbf{\textcolor{green}{#1 -- LL}}}}

	\newcommand{\BMc}[1]{{\textbf{\textcolor{green}{#1 -- BM}}}}

	\newcommand{\SRMout}[1]{{\textbf{\textcolor{purple}{\sout{#1}}}}}
	\newcommand{\SRMc}[1]{{\textbf{\textcolor{purple}{[#1 -- Sam]}}}}

}{%
	\newcommand{\BCcom}[1]{}
	\newcommand{\BCout}[1]{}

	\newcommand{\JSout}[1]{}
	\newcommand{\JSeqout}[1]{}
	\newcommand{\JSc}[1]{}

	\newcommand{\LLc}[1]{}

	\newcommand{\BMc}[1]{}

	\newcommand{\SRMout}[1]{}
	\newcommand{\SRMc}[1]{}

}

\begin{document}

\preprint{SITP/16-15}

\title{Equivalent Equations of Motion\\
for Gravity and Entropy}% Force line breaks with \\

\author{Bart{\l}omiej Czech}
  \email{czech@stanford.edu}
\author{Lampros Lamprou}%
 \email{llamprou@stanford.edu}
\author{Samuel McCandlish}%
 \email{smccandlish@stanford.edu}
\author{Benjamin Mosk}%
\email{bmosk1@stanford.edu}
\affiliation{%
  Stanford Institute for Theoretical Physics, Department of Physics, Stanford University\\
	Stanford, CA 94305, USA
}%

\author{James Sully}
\email{jsully@slac.stanford.edu}
\affiliation{
 Theory Group, SLAC National Accelerator Laboratory\\ Menlo Park,
	CA 94025, USA
}%
%\affiliation{

\date{\today}% It is always \today, today,

\begin{abstract}
We demonstrate an equivalence between the wave equation obeyed by the entanglement entropy of CFT subregions and the linearized bulk Einstein equation in Anti-de Sitter pace.  
In doing so, we make use of the formalism of kinematic space \cite{Czech:2015qta} and fields on this space, introduced in \cite{Czech:2016xec}.
We show that the gravitational dynamics are equivalent to a gauge invariant wave-equation on kinematic space and that this equation arises in natural correspondence to the conformal Casimir equation in the CFT.
\end{abstract}

\maketitle

\section{\label{sec:introduction}Introduction}

The recent direct detection of gravitational waves \cite{Abbott:2016blz}
adds to an impressive list of observational tests of general relativity,
the theory that describes long-distance dynamics of spacetime. At
the quantum level, however, the fate of spacetime requires a more careful
assessment; indeed, it is not even clear what the fundamental degrees
of freedom in a theory of quantum gravity are. The holographic principle
\cite{Susskind:1994vu,Hooft:1993gx} suggests that these degrees of freedom should in fact be nonlocal. 

This notion is made explicit in the AdS/CFT duality, an equivalence
between $d-$dimensional conformal field theories (CFTs) and $(d+1)$-dimensional
gravitational systems with Anti de Sitter (AdS) asymptotics. 
In particular, the proposal of Ryu and Takayangi (RT) \cite{Ryu:2006bv,Hubeny:2007xt} relates the
entanglement entropy $S$ of a CFT region $B$ to the area of a
bulk extremal surface $\tilde{B}$,
\begin{equation}
S\left(B\right)=\min_{\partial\tilde{B}=\partial B}\frac{{\rm area}\,(\tilde{B})}{4G_{N}}.\label{eq:ryu-takayanagi}
\end{equation}
Developments over the past few years \cite{Lashkari:2013koa,Faulkner:2013ica,Swingle:2014uza} \cite{Nozaki:2013vta,Bhattacharya:2013bna} have shown that the dynamics of this nonlocal CFT quantity is closely related to the bulk Einstein equation.

In recent work \cite{Czech:2016xec} we demonstrated an extension
of the RT proposal to other bulk scalar fields. We introduced the \emph{OPE
block}, the contribution to the CFT operator product expansion (OPE)
from a single conformal family, and equated it to the \emph{Radon
transform}, the integral of a bulk scalar field over a minimal surface.
Both of these objects obey an equation of motion in \emph{kinematic
space}, a space which geometrizes the set of bulk surfaces of a given
dimension. We showed that this kinematic equation of motion emerges
directly from the bulk Klein-Gordon equation.

In this paper, we will demonstrate the relationship between the Ryu-Takayanagi
proposal and our results on kinematic space. Specifically, we will
show that the wave equations obeyed by a perturbation to entanglement entropy \cite{Wong:2013aa,Nozaki:2013vta,Bhattacharya:2013bna,PhysRevLett.116.061602}  correspond directly to the linearized bulk Einstein equation with
matter via an \emph{intertwining} relation of the Radon transform:
\begin{align*}
\left(\square_{\mathcal{K}}+2d\right)\delta S=0\quad\leftrightarrow\quad & \text{Einstein equations}\\
\left(\square_{\rm{dS}}+d\right)\delta S=0\quad\leftrightarrow\quad & \text{Hamiltonian constrant} 
\end{align*}
where the Laplacians $\square_\mathcal{K}$ and $\square_{\rm{dS}}$ on kinematic space arise from conformal Casimir equations. This clarifies and connects the results of \cite{Lashkari:2013koa,Faulkner:2013ica,Swingle:2014uza,Nozaki:2013vta,Bhattacharya:2013bna}.

Before stating our result explicitly, we begin by outlining the results
of \cite{Czech:2016xec} on kinematic space, the OPE block, and the
Radon transform.

\section{\label{sec:kinematicspace}Scalar Kinematic Dictionary}

The kinematic dictionary, introduced in \cite{Czech:2016xec} and
recently also explored in \cite{deBoer:2016pqk}, connects Radon transforms
of AdS-fields with OPE-blocks in the dual CFT. Here we outline the
basic formalism.

\paragraph{Radon transform}

The Radon transform is a map from functions $f\left(x\right)$ on
some manifold to functions on the space of $n$-dimensional totally
geodesic submanifolds $\tilde{B}$. It is defined via the integral
transform
\begin{equation}
R\left[f\right] \left(\tilde{B}\right)=\int_{\tilde{B}}dA\,f\label{eq:radon-transform}
\end{equation}
where $dA$ is the induced area element on the surface $\tilde{B}$.
Though \cite{Czech:2016xec} also considered the case $n=1$ of geodesics,
we will focus here on the case $n=d-1$ of codimension-2 minimal surfaces
in ${\rm AdS}_{d+1}$. 

\begin{figure}
\begin{centering}
\includegraphics[width=1\columnwidth]{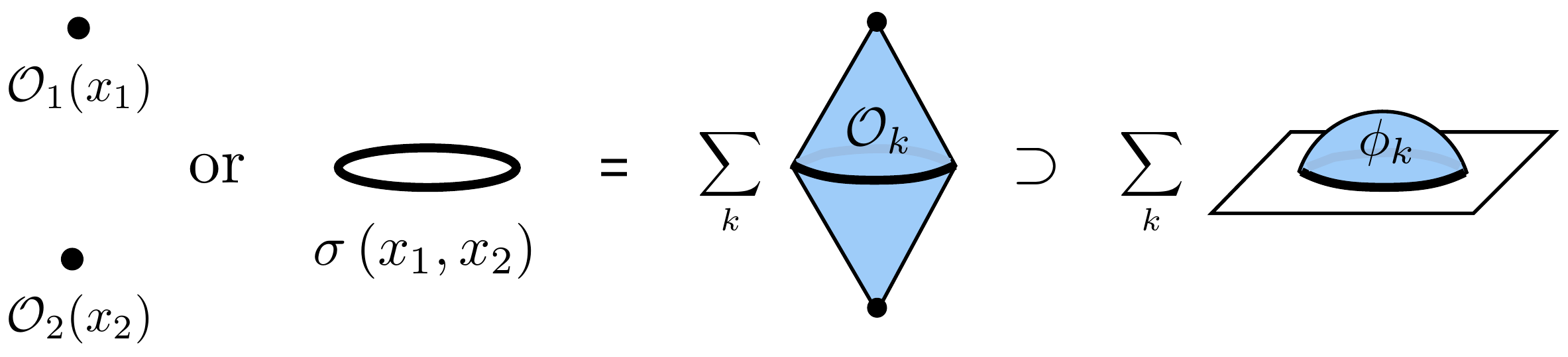} 
\par\end{centering}
\caption{\label{fig:opemin}A point in kinematic space labels two timelike-separated boundary points, $x_1,x_2$, or equivalently a codimension-2 sphere formed by their causal domain. Operators $\sigma(x_1,x_2)$ localized on the sphere have an OPE expansion in terms of OPE blocks for primaries $\mathcal{O}_k$, where contributions can be identified as bulk surface operators $\phi_k$. Figure from \cite{Czech:2016xec}.}
\end{figure}

It is useful to define an auxiliary space $\mathcal{K}$, which we
call \emph{kinematic space}, to organize information about bulk surfaces.
A point in $\mathcal{K}$ denotes equivalently any of the following:
\begin{itemize}
\item a particular bulk minimal surface, $\tilde B$,
\item the boundary sphere where that surface ends, $B$,
\item the two timelike separated boundary points $x_{1},x_{2}$ at the tips
of the causal development of this sphere (see Fig. \ref{fig:opemin}), $\blacklozenge_{12}$.
\end{itemize}
The points of $\mathcal{K}$ are most conveniently parameterized by
the two points $\left(x_{1},x_{2}\right)$. 
When the context is clear, we will often denote any of the above three objects by the pair $\left(x_{1},x_{2}\right)$.
The conformal group ${\rm SO}\left(d,2\right)$
then endows kinematic space with a metric structure of $\left(d,d\right)$
signature (see \cite{Czech:2016xec}),
\begin{equation}
ds^{2}=\frac{I_{\mu\nu}\left(x_{1}-x_{2}\right)}{\left(\frac{x_{1}-x_{2}}{2}\right)^{2}}dx_{1}^{\mu}dx_{2}^{\nu}
\end{equation}
where $I_{\mu\nu}\left(x\right)=\eta_{\mu\nu}-2\frac{x_{\mu}x_{\nu}}{x^{2}}$
is the CFT inversion tensor. 

\begin{figure}
	\centering
		\includegraphics[width=0.8\columnwidth]{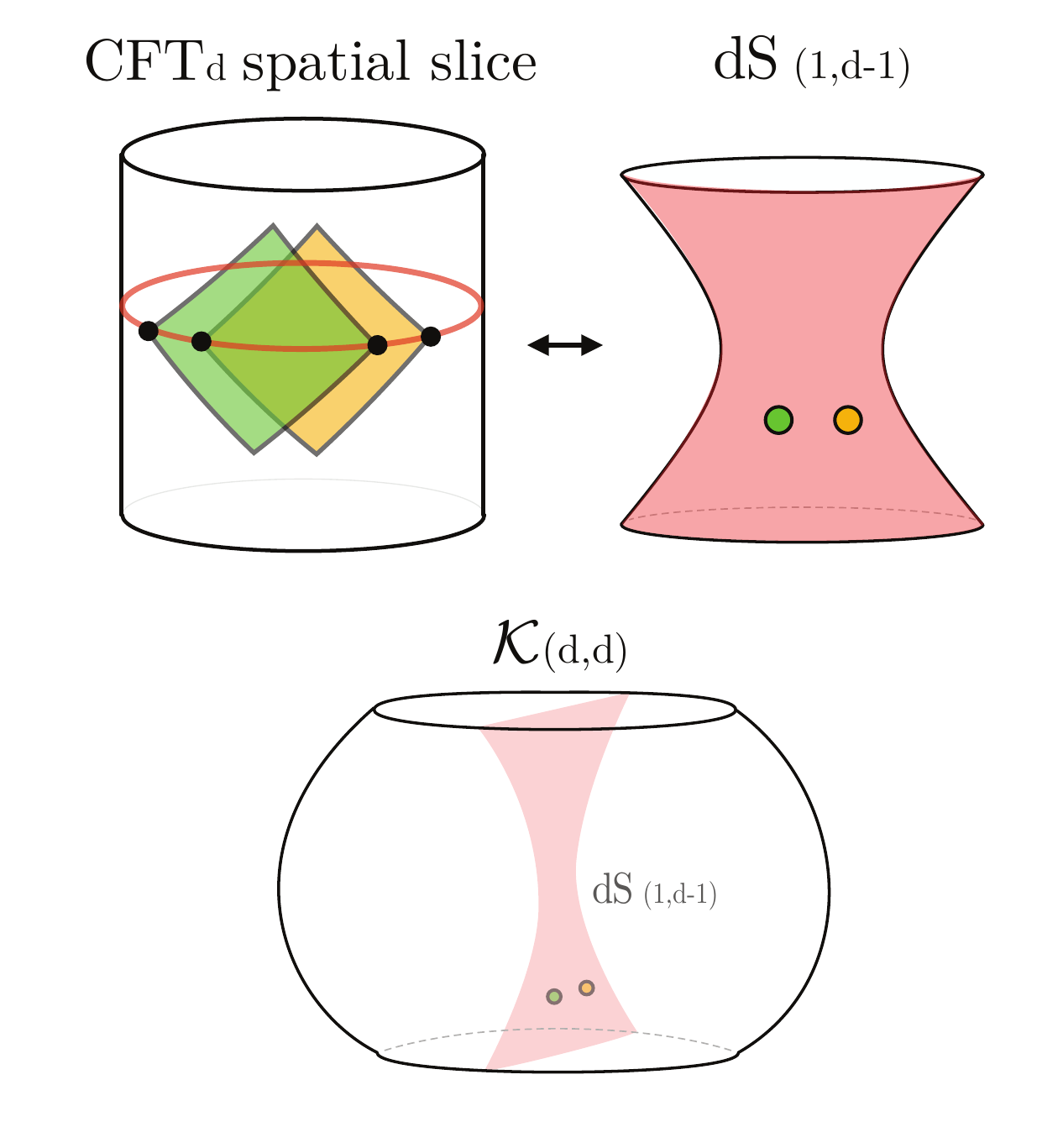}
		\caption{The kinematic space for spherical regions that lie on a single time slice is given by $d$-dimensional Lorentzian de Sitter space. The de Sitter space is a corresponding slice of the larger kinematic space for all (boosted) spherical regions, which is $2d$-dimensional with signature $(d,d)$.}
	\label{fig:slice}
\end{figure}
It will also be useful to note that if we consider only spheres living
in a particular equal-time slice of ${\rm AdS}_{d+1}$, preserved
by an ${\rm SO}\left(d,1\right)$ subgroup, the corresponding slice
of $\mathcal{K}$ has the structure of a $d$-dimensional de Sitter
(dS) space \cite{solanes,Czech:2015qta,Czech:2016xec,PhysRevLett.116.061602} (see Fig. \ref{fig:slice}). There is one such dS slice for each time slice
of AdS. For instance, if we consider the $t=0$ slice of AdS and parameterize
the spheres it contains by their radius $R$ and center $\vec{x}$,
the induced metric on this slice is given by
\begin{equation}
ds^{2}=\frac{-dR^{2}+d\vec{x}^{2}}{R^{2}}.
\end{equation}
A particularly useful feature of the kinematic space $\mathcal{K}$ is that the domain of the CFT, including the time direction, appears as a spacelike surface at $x_1 = x_2$.  This allows us to impose boundary conditions and choose a causal propagator in the usual way when solving wave equations in kinematic space.

\paragraph{OPE-blocks}

We now introduce a CFT object whose domain is also $\mathcal{K}$.
Recall that in a CFT, the product of two identical scalar local operators
can be expanded in terms of the global primary operators of the theory
as
\begin{equation}
\frac{\mathcal{O}\left(x_{1}\right)\mathcal{O}\left(x_{2}\right)}{\left\langle \mathcal{O}\left(x_{1}\right)\mathcal{O}\left(x_{2}\right)\right\rangle }=\sum_k C_{\mathcal{O}\mathcal{O}k}\underbrace{{\scriptstyle \left|x_{12}\right|^{\Delta_{k}}\left(1+b_{1}x_{12}^{\mu}\partial_{\mu}+\cdots\right)}\mathcal{O}_{k}\left(x_{2}\right)}_{{\textstyle \mathcal{B}_{k}\left(x_{1},x_{2}\right)}}\label{eq:ope-block-definition}
\end{equation}
where $x_{12}^{\mu}=x_{1}^{\mu}-x_{2}^{\mu}$. The coefficients $C_{\mathcal{O}\mathcal{O}k}$
are known as the OPE coefficients and are theory-dependent, while
the coefficients $b_{i}$ depend only on the scaling dimension $\Delta_{k}$
of the primary operator $\mathcal{O}_{k}$. We have grouped the contributions
from a single primary to the OPE into an object $\mathcal{B}_{k}\left(x_{1},x_{2}\right)$,
which we call the OPE-block. In \cite{Czech:2016xec}, we showed that
operators $\sigma\left(x_{1},x_{2}\right)$ localized on a CFT sphere
can be expanded in terms of the same OPE blocks as
\begin{equation}
\frac{\sigma\left(x_{1},x_{2}\right)}{\left\langle \sigma\left(x_{1},x_{2}\right)\right\rangle }=\sum_k C_{\sigma k}\mathcal{B}_{k}\left(x_{1},x_{2}\right)\label{eq:sphere-operator-blocks}
\end{equation}
where $C_{\sigma k}$ are theory-dependent ``OPE'' coefficients.

The OPE-block also has a compact integral expression \cite{Czech:2016xec} over the causal diamond formed by the points $x_1,x_2$. For a scalar, this is just:
\begin{equation}
\mathcal{B}_k(x_1,x_2) = N_k \int_{\blacklozenge_{12}} d^d x_3 \left(\frac{x_{13}x_{23}}{x_{12}}\right)^{\Delta_k-d} \mathcal{O}_k(x_3) \, .
\label{eq:smearing}
\end{equation}
OPE-blocks are useful as CFT objects in their own right; however,
they become even more powerful in the presence of an AdS dual, as
we describe next.

\paragraph{Kinematic Dictionary}

At leading order in the $N\rightarrow\infty$ limit, the OPE-block
$\mathcal{B}_{k}\left(x_{1},x_{2}\right)$ and the Radon transform
$R\left[\phi_{k}\right]\left(x_{1},x_{2}\right)$ of the dual AdS field are directly
related:
\begin{equation}
\mathcal{B}_{k}\left(x_{1},x_{2}\right)=\frac{1}{c_{\Delta_{k}}}R\left[\phi_{k}\right]\left(x_{1},x_{2}\right)\label{eq:kinematicdictionary}
\end{equation}
where $c_{\Delta}$ is a constant depending only on the dimension
$d$ and the scaling dimension $\Delta_{k}$. To prove this, in \cite{Czech:2016xec} we showed
that both sides obey the same equation of motion with the same boundary
conditions:
\begin{align}
\left(\square_{\mathcal{K}}+m_{k}^{2}\right)\mathcal{B}_{k}=0 & \qquad\left(\square_{\mathcal{K}}+m_{k}^{2}\right)R\left[\phi_{k}\right]=0\nonumber \\
\mathcal{B}_{k}\sim\left|x_{12}\right|^{\Delta_{k}}\mathcal{O}_{k} & \qquad R\left[\phi_k\right]\sim c_{\Delta_{k}}\left|x_{12}\right|^{\Delta_{k}}\mathcal{O}_{k} \, .\label{eq:kinematic-dictionary-matching}
\end{align}

The OPE block equation of motion comes from a conformal Casimir equation,
\begin{equation}
\left[L^{2},\mathcal{B}_{k}\right]=\square_{\mathcal{K}}\mathcal{B}_{k}=C_{k}\mathcal{B}_{k}\label{eq:casimir-representation}
\end{equation}
where $L^{2}$ is the Casimir element of the conformal group ${\rm SO}\left(d,2\right)$,
with eigenvalue 
\begin{equation}
C^{{\rm SO}\left(d,2\right)}_k=-\Delta\left(\Delta-d\right)-\ell\left(\ell+d-2\right)=-m^{2}_k.\label{eq:casimir-eigenvalue}
\end{equation}
The Casimir element is represented on kinematic space scalar fields
by the Laplacian $\square_{\mathcal{K}}$, yielding the equation of
motion. The boundary condition as $x_{1}\rightarrow x_{2}$ comes
from inspecting the definition \ref{eq:ope-block-definition}.

To find the equation of motion for the Radon transform we use an \emph{intertwining}
property:
\begin{equation}
\square_{\mathcal{K}}R\phi=-R\square_{{\rm AdS}}\phi.\label{eq:scalar-intertwinement}
\end{equation}
Together with the Klein-Gordon equation, this implies an equation
of motion for the Radon transform:
\begin{equation}
\square_{{\rm AdS}}\phi_{k}=m_{k}^{2}\phi_{k}\quad\implies\quad\square_{\mathcal{K}}\phi_{k}=-m_{k}^{2}\phi_{k}.\label{eq:scalar-intertwinement-eom}
\end{equation}
The boundary condition then comes from the AdS/CFT dictionary $\phi_{k}\left(x,z\right)\rightarrow z^{\Delta_{k}}\mathcal{O}_{k}\left(x\right)$.

Since kinematic space has signature $\left(d,d\right)$, an additional
$d-1$ equations are required to fix a solution uniquely. These take
the form of constraint equations, explained in detail in \cite{Czech:2016xec,deBoer:2016pqk}.
These equations of motion together with the boundary conditions establish
the validity of \ref{eq:kinematicdictionary}.

In the remainder of this paper, we will use the same techniques to
extend the kinematic dictionary to the CFT stress tensor, which is
dual to the bulk metric perturbation. We will extend this dictionary
to first subleading order in the $1/N$ expansion, finding that the
correction is precisely that found in \cite{Faulkner:2013ana}. This will allow us
to prove an equivalence between the linearized Einstein's equations
in the bulk and a simple equation satisfied by the stress tensor OPE-block,
which we show is equal to the modular Hamiltonian.

\section{Tensor Radon Transforms and Einstein's Equations\label{sec:einstein-and-area-equations}}

In this section, we will show that the linearized Einstein equations
are equivalent to a set of equations obeyed by the fluctuation in the
area of the minimal surfaces. We will do this in a similar way as
in Eq. \ref{eq:scalar-intertwinement-eom}, by using an intertwining
relation. The goal will be to find an analog of the field equation
and boundary conditions of Eq. \ref{eq:kinematic-dictionary-matching};
we will then match to CFT quantities in the following section.

Since the bulk field of interest, the metric perturbation $\delta g_{\mu\nu}$,
is a tensor, we must first introduce a tensor analog of the Radon
transform Eq. \ref{eq:radon-transform}. For a symmetric 2-tensor field
$s_{\mu\nu}$, we define the \emph{longitudinal }and \emph{transverse}
Radon transforms, denoted $R_{\parallel}$ and $R_{\perp}$ respectively,
as
\begin{align}
R_{\parallel}\left[s_{\mu\nu}\right]\left(x_1,x_2\right) & =\int_{\tilde{B}_{12}}dA\,h^{\mu\nu}s_{\mu\nu}\nonumber\\
R_{\perp}\left[s_{\mu\nu}\right]\left(x_1,x_2\right) & =\int_{\tilde{B}_{12}}dA\,\left(g^{\mu\nu}-h^{\mu\nu}\right)s_{\mu\nu}.
\end{align}
Here, $h_{\mu\nu}$ denotes the induced metric on the surface $\tilde{B}_{12}$.
As before, these transforms output a scalar function on kinematic
space; we write indices on the left side only to indicate that the
input is a tensor. 

We now note some useful identities for the tensor Radon transform. First, note that the
sum of the two tensor transforms is just a scalar Radon transform
of ${\rm tr}s$. This implies that the two are related by a trace-reversal
of the input tensor,
\begin{align}
R_{\parallel}\left[s_{\mu\nu}\right] & =-R_{\perp}\left[s_{\mu\nu}-\frac{1}{2}g_{\mu\nu}{\rm tr}s\right]\nonumber \\
R_{\perp}\left[s_{\mu\nu}\right] & =-R_{\parallel}\left[s_{\mu\nu}-\frac{1}{d-1}g_{\mu\nu}{\rm tr}s\right].\label{eq:tensor-transform-relationship}
\end{align}

Before we continue, let us pause to note a striking simplification
of the full nonlinear Einstein equation when written in terms of tensor
Radon transforms. The Einstein equation takes the form

\begin{equation}
R_{\mu\nu}-\frac{1}{2}g_{\mu\nu}R= 8 \pi G_N T_{\mu\nu}.
\end{equation}
where we have defined $T_{\mu\nu}$ to include the cosmological constant
term. If we apply the transverse transform $R_{\perp}$ to both sides,
using the identity \ref{eq:tensor-transform-relationship}, we obtain
\begin{equation}
\tfrac{1}{4G_N}R_{\parallel}\left[R_{\mu\nu}\right]+2\pi R_{\perp}\left[T_{\mu\nu}\right]=0.\label{eq:integrated-einstein}
\end{equation}
This remarkable simplification occurs only when we integrate over
a codimension-2 surface, due to the appearance of the coefficient
$\frac{1}{2}$ in the Einstein tensor\footnote{However, note for a dimension-2 (rather than codimension-2) surface,
the longitudinal transform of the Einstein equations yields the equation
\begin{equation}
R^{(2)}_{\perp}\left[R_{\mu\nu}\right]+R^{(2)}_{\parallel}\left[T_{\mu\nu}\right]=0.
\end{equation}
}.

Let us now consider the linearized version of Eq. \ref{eq:integrated-einstein}.
Setting $T_{\mu\nu}=-\frac{\Lambda}{8 \pi G_N} g_{\mu\nu}+ \delta T_{\mu\nu}$, with
$\Lambda=-\frac{1}{2}d\left(d-1\right)$ for ${\rm AdS}_{d+1}$, we
have
\begin{equation}
\tfrac{1}{4G_N} R_{\parallel}\left[\delta R_{\mu\nu}\right]=- 2\pi \, R_{\perp}\left[\delta T_{\mu\nu}\right]-\tfrac{2d}{4G_N}\delta A\label{eq:linearized-integrated-einstein}
\end{equation}
where $\delta A$ is the first order change in the area of the surface
of integration. In terms of the tensor Radon transform, the area perturbation
can be written as 
\begin{equation}
\delta A\left(x_1,x_2\right)=\frac{1}{2}\int_{\tilde{B}_{12}}h^{\mu\nu}\delta g_{\mu\nu}dA=\frac{1}{2}R_{\parallel}\delta g.\label{eq:area-perturbation-as-a-transform}
\end{equation}
We would like to specialize to minimal surfaces in ${\rm AdS}_{d+1}$,
and recast the linearized equation \ref{eq:linearized-integrated-einstein}
as an equation of motion in kinematic space. To do this, we make use
of an intertwining relation analogous to \ref{eq:scalar-intertwinement},
\begin{equation}
\square_{K}R_{\parallel}\left[s_{\mu\nu}\right]=-R_{\parallel}\left[\left(\nabla^{2}+2\left(d+1\right)\right)s_{\mu\nu}-2g_{\mu\nu}{\rm tr}s\right],\label{eq:tensor-intertwinement}
\end{equation}
where $\nabla^{2}=\nabla_{\alpha}\nabla^{\alpha}$ denotes the covariant
Laplacian. The right side of this equation is given by the action
of the casimir $L_{{\rm SO}\left(d,2\right)}^{2}$ on $s_{\mu\nu}$,
as shown in Appendix \ref{Tensor_Int}. In fact, for the case of $\delta g_{\mu\nu}$,
Eq. \ref{eq:tensor-intertwinement} can be rewritten as
\begin{equation}
\square_{K}\delta A=R_{\parallel}\left[\delta R_{\mu\nu}\right],\label{eq:einstein-tensor-intertwinement-full}
\end{equation}
where $\delta R_{\mu\nu}$ is the variation of the Ricci tensor due
to the variation $\delta g_{\mu\nu}$ in the metric; this is shown
explicitly in Appendix \ref{Einst_Eq}. Together with \ref{eq:linearized-integrated-einstein},
this implies the equation of motion
\begin{equation}
\left(\square_{K}+2d\right)\frac{\delta A}{4 G_N} =-2 \pi \, R_{\perp}\left[\delta T\right].\label{eq:area-perturbation-ks-eqn}
\end{equation}
We have thus shown that the area perturbation $\delta A$ obeys an
equation of motion in kinematic space as a consequence of the linearized Einstein equation
about AdS. 

To show complete \emph{equivalence} between the kinematic equation of motion and Einstein equations, it remains only to show that the tensor Radon transform is invertible (up to diffeomorphisms). 
Unfortunately, while reasonable, we are not aware of a proof of this fact in the literature, and our statement of equivalence must carry a technical asterisk awaiting further input from the mathematical community.

To avoid the technical problem in the preceding paragraph, we will now prove an additional equation of motion for the area perturbation,
but this time restricting ourselves to surfaces on a time slice of
AdS; this corresponds to a particular de Sitter slice of kinematic
space. Using the same techniques as above, we prove in Appendix \ref{Einst_Eq}
that 
\begin{equation}
\left(\square_{{\rm dS}}+d\right)\delta A=R\left[\delta\left(R_{\mu\nu}-\frac{1}{2}Rg_{\mu\nu}\right)\hat{t}^{\mu}\hat{t}^{\nu}\right].\label{eq:einstein-tensor-intertwinement-slice}
\end{equation}
The right hand side is a $tt$ component of the Einstein tensor; hence,
using the Einstein equation, we find
\begin{equation}
\left(\square_{{\rm dS}}+d\right)\frac{\delta A}{4 G_N}= 2  \pi R\left[\delta T_{00}\right].\label{eq:area-perturbation-ds-eqn}
\end{equation}
Here, $T_{00}$ denotes the energy density relative to the particular
AdS time slice we are considering; there is a separate de Sitter equation
for each time slice. Hence, the Hamiltonian constraint of the Einstein
equation implies a de Sitter equation of motion for the area perturbation.
In thise case, since the scalar Radon transform is known to be injective \cite{Helgason1999}, Eq. \ref{eq:area-perturbation-ds-eqn}
is \emph{equivalent} to the Hamiltonian constraint on a time slice,
and the collection of de Sitter equations for every slice is equivalent
to the full linearized Einstein equation, because knowing $E_{00} = T_{00}$ for every choice of $\hat t $ implies $E_{\mu \nu} = T_{\mu \nu}$.

To complete the description of the Cauchy problem for the area perturbation
$\delta A$ in kinematic space, we must fix boundary conditions. This
can be done using the extrapolate dictionary for the metric perturbation \cite{Balasubramanian:1999re,Myers:1999psa,deHaro:2000vlm},
and was shown in \cite{Faulkner:2013ica} to be 
\begin{equation}
\delta A\left(x_{0},R\right)\underset{R\rightarrow0}{\sim}R^{d}T_{00}\left(x_0\right) \frac{8\pi G_{N}\Omega_{d-2}}{d^{2}-1}.\label{eq:area-boundary-condition}
\end{equation}

Now that we have formulated the Cauchy problem for the area perturbation
in the form of Eqn. \ref{eq:area-boundary-condition} along with either
\ref{eq:area-perturbation-ks-eqn} or \ref{eq:area-perturbation-ds-eqn},
we can proceed to match with CFT variables.

\section{Modular Hamiltonian and the Tensor Kinematic Dictionary}

In this section we use the Radon-transformed Einstein equations we have just derived to give a novel derivation of the quantum-corrected Ryu-Takayanagi formula. 

To begin, we take a moment to review the entanglement first law and the
form of the vacuum modular Hamiltonian. Given a quantum mechanical
system in a certain state $|\psi\rangle$, the state of a subsystem
$B$ is described by the reduced density matrix $\rho_{B}$, obtained
from $|\psi\rangle\langle\psi|$ by tracing over the degrees of freedom
of the complement $B^{c}$. For such a subsystem, the entanglement
entropy is defined as: 
\begin{equation}
S=-{\rm tr}\rho_{B}\log\rho_{B}.\label{eq:vn}
\end{equation}
The \emph{modular Hamiltonian} $H_{{\rm mod}}$ of the state $\rho_{B}$
is then defined implicitly by
\begin{equation}
\rho_{B}=\frac{e^{-H_{{\rm mod}}}}{{\rm tr}\left(e^{-H_{{\rm mod}}}\right)}.
\end{equation}
Using this expression, the change in the entanglement entropy of $A$
due to a small perturbation of the state can be compactly expressed
as: 
\begin{equation}
\delta S=\delta\left\langle H_{\text{mod}}\right\rangle \, . \label{1stlaw}
\end{equation}
This equation is known as the first law of entanglement entropy.

When $B$ is a ball of radius $R$ in the vaccum state of a CFT, the modular
Hamiltonian can be written as 
\begin{equation}
H_{\text{mod}}=2\pi\int\limits _{B}d^{d-1}x\frac{R^{2}-\left(x-x_{0}\right)^{2}}{2R}T_{00}\left(x\right) \, ,\label{eq:Hmod}
\end{equation}
where $T_{00}$ is the energy density in the CFT. The form of the
vacuum modular Hamiltonian was computed in \cite{Casini2011}. The
fact that $H_{{\rm mod}}$ is an OPE block was pointed out in \cite{Czech:2016xec}; in Appendix \ref{ModHam} we give the details for
general dimension.

We can now make contact with the equations of motion (\ref{eq:area-perturbation-ks-eqn},
\ref{eq:area-perturbation-ds-eqn}). It was pointed out by \cite{PhysRevLett.116.061602}
that the vacuum modular Hamiltonian, when viewed as a field on kinematic
space, obeys a de Sitter wave equation
\begin{equation}
\left(\square_{{\rm dS}}+d\right)H_{{\rm mod}}=0.\label{eq:hmod-ds-eom}
\end{equation}
It in fact obeys a separate equation for each CFT time slice, each
of which has a corresponding de Sitter slice of the full kinematic
space $\mathcal{K}$. To see this, note that the ${\rm SO}\left(d,1\right)$
subgroup of the conformal group that preserves a time slice has the
Casimir
\begin{equation}
C_{{\rm SO}\left(d,1\right)}=-\Delta\left(\Delta-d+1\right)-\ell\left(\ell+d-3\right).
\end{equation}
Since $T_{00}\left(x\right)$ transforms as a scalar of dimension
$\Delta=d$ under this subgroup, its Casimir eigenvalue is $-d$.
Then $H_{{\rm mod}}$, being an integral of $T_{00}$, satisfies $\left[L_{{\rm SO}\left(d,1\right)}^{2},H_{{\rm mod}}\right]=-dH_{{\rm mod}}$.
Since $H_{{\rm mod}}$ transforms as a scalar field on kinematic space,
the Casimir $L_{{\rm SO}\left(d,1\right)}^{2}$ is represented by
the Laplacian $\square_{{\rm dS}}$, yielding the equation \ref{eq:hmod-ds-eom}.
It follows similarly that $H_{{\rm mod}}$ obeys an equation of motion
on the full kinematic space, 
\begin{equation}
\left(\square_{\mathcal{K}}+2d\right)H_{{\rm mod}}=0,\label{eq:hmod-full-eom}
\end{equation}
with eigenvalue $2d$ coming from Eq. \ref{eq:casimir-eigenvalue}
\cite{Czech:2016xec}.

Having written an equation of motion for $H_{{\rm mod}}$, we would
now like to check the boundary conditions in kinematic space for $H_{{\rm mod}}$.
%\SRMc{Maybe say something about restricting to the $t=0$ slice, then generalizing.} 
Taking the limit $R\rightarrow0$ of Eq. \ref{eq:Hmod}, we find 
\begin{align}
H_{\text{mod}}\left(x_0,R\right) & \underset{R\rightarrow0}{\sim}R^{d}T_{00}\left(x_0\right)\frac{2\pi\Omega_{d-2}}{d^{2}-1}.\label{eq:hmod-boundary-conditions}
\end{align}
We can now compare directly with the results of Sec. \ref{sec:einstein-and-area-equations}.
First, let us consider the leading order in $1/N$ behavior, for which
$\delta T_{\mu\nu}=0$. In that case, Eqns. \ref{eq:area-perturbation-ds-eqn}
and \ref{eq:hmod-ds-eom} match, and the boundary conditions \ref{eq:area-boundary-condition}
and \ref{eq:hmod-boundary-conditions} differ only by a constant $4G_{N}$.
This gives us the leading-order kinematic dictionary:
\begin{equation}
H_{{\rm mod}}=\frac{\delta A}{4G_{N}}+O\left(N^{0}\right).\label{eq:leading-order-rt}
\end{equation}
Of course, this is just the linearized Ryu-Takayanagi formula \ref{eq:ryu-takayanagi}.

\begin{figure}
	\centering
		\includegraphics[width=1\columnwidth]{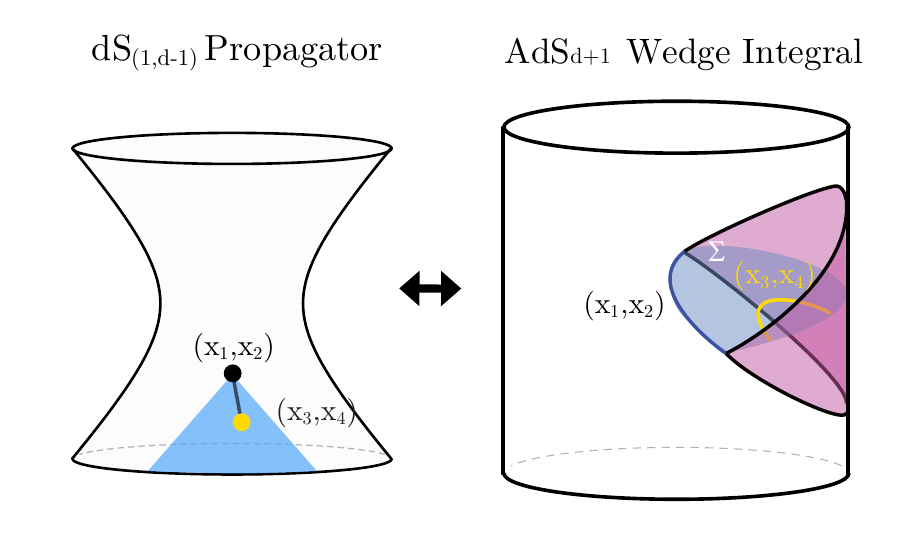}
	\caption{The retarded de Sitter propagator corresponds to an integral over all the geodesics that lie on the $t=0$ slice of the causal wedge. This integral reconstructs the bulk modular Hamiltonian, which is also an integral over the same spatial slice.}
	\label{fig:HmodProp}
\end{figure}

To find the $O\left(N^{0}\right)$ correction to the dictionary, we
must find an object $X$ which satisfies 
\begin{equation}
\left(\square_{dS}+d\right)X= -2 \pi R\left[\delta T_{00}\right].\label{eq:correction-equation}
\end{equation}
Using Eq. \ref{eq:area-perturbation-ds-eqn}, this will the guarantee that $\frac{\delta A}{4G_{N}}+X$ satisfies
the same EOM as $H_{{\rm mod}}$, Eq. \ref{eq:hmod-ds-eom}. The solution can
be written as
\begin{multline}
X\left(x_1,x_2\right)=\\-2 \pi \int_{\triangle} G_{{\rm dS}}^{{\rm ret}}\left(x_1,x_2;x_3,x_4\right)R\left[\delta T_{00}\right]\left(x_3,x_4\right) dV
\end{multline}
where the integration region is the past light cone of $\left(x_1,x_2\right)$,
and where $G_{{\rm dS}}$ is a bulk-to-bulk kinematic space causal
propagator. The result is
\begin{equation}
X= 2 \pi \int_{\Sigma}T_{\mu\nu}\xi^{\mu}d\Sigma^{\nu}=H_{{\rm bulk}} \label{eq:bulk-modular-hamiltonian}
\end{equation}
where $\xi^{\mu}$ is the Killing vector corresponding to modular
flow \cite{deBoer:2016pqk}, and where $d\Sigma^{\nu}$ is the timelike unit normal vector
to $\Sigma$ (see Figure \ref{fig:HmodProp}). 
%\SRMc{Say why boundary conditions are not affected.}
We immediately recognize that $X$ is none other than $H_{{\rm bulk}}$,
the bulk vacuum modular Hamiltonian for a Rindler wedge. (In Appendix \ref{sec:wedge-intertwinement}, we show
via an intertwining relation that $H_{{\rm bulk}}$ indeed satisfies
Eqn. \ref{eq:correction-equation}.) 

We thus arrive at the corrected
kinematic dictionary, 
\begin{equation}
H_{{\rm mod}}=\frac{\delta A}{4G_{N}}+H_{{\rm bulk}},
\label{eq:full-kinematic-dictionary}
\end{equation}
which matches the FLM correction to the Ryu-Takayanagi formula \cite{Faulkner:2013ana,Jafferis:2015aa}.
We hence recognize the Ryu-Takayanagi formula arises as a special case of
our more general kinematic dictionary.

%\begin{figure}
%\includegraphics[width=1\columnwidth]{entangled-trio}\caption{Any two of these three facts imply the third, at a linearized level.}
%\end{figure}

\section{Discussion}

We have given a simple and elegant demonstration of the equivalence of the Einstein equations and the quantum-corrected Ryu-Takayanagi formula for linearized perturbations about the vacuum. 
In doing so, we exploited the fact that both the bulk Einstein equations and the boundary modular Hamiltonian obey simple dynamical equations in an auxiliary kinematic space. 

The derivation of the quantum corrected Ryu-Takayanagi formula from the Einstein equations was already described in \cite{Faulkner:2013ana}. 
While the previous work is more general, the bulk quantum contribution depends on a generally non-local and unknown modular Hamiltonian. 
Our approach, on the other hand, makes more explicit how quantum corrections arise from bulk interactions. 
We hope that these techniques will prove insightful when extended away from simple regions of the vacuum state. 

This paper was also primarily focused on demonstrating how the bulk Einstein equations imply the Ryu-Takayanagi formula, but our work equally leads to the reverse statement. 
This is most mathematically rigorous when we make use of the kinematic space for a single-time slice and the scalar Radon transform to derive the $tt$-component of the Einstein equations (all components then follow by appropriate boosts, as in \cite{Lashkari:2013koa}).
Here, we can derive the local bulk EOM because the scalar Radon transform on Hyperbolic space has been proven to be invertible \cite{Helgason1999}.
However, to exploit the full kinematic space and directly derive any component of the Einstein equations, we must invert the tensor Radon transform. 
While the invertibility of these transforms (up to diffeomorphism) is well-motivated, we are not aware of a mathematical proof. 
Thus, a stickler for rigor will conclude that only an integrated version of the Einstien equations has been derived by this second approach.

As the gravitational equations of motion intertwine to become a kinematic equation of motion which is fixed by conformal invariance, it may be confusing to wonder what happens to the gravitational equations of motion in a generalized theory of gravity. 
However, the particular gravitational equation of motion is determined not by the fixed kinematic EOM, but by the choice of entropy functional. 
In particular, it was shown in \cite{Lashkari:2013koa,Faulkner:2013ica,Swingle:2014uza} using the
Wald-Iyer formalism \cite{Iyer:1995kg} that the entanglement first law \ref{1stlaw}
can be written for perturbations of the vacuum as 
\begin{equation}
0=\delta S-\delta\left\langle H_{{\rm mod}}\right\rangle = 2\pi \int_{\Sigma}\left[\delta E_{\mu\nu}-\delta T_{\mu\nu}\right]\xi^{\mu}d\Sigma^{\nu} \, .\label{eq:mark-wald}
\end{equation}
In this equation, $\delta E_{\mu\nu}$ is the linearized equation of motion for a general theory of gravity, and the point-wise equation of motion follows from considering all surfaces $\Sigma$.

In upcoming work, one of us \cite{Ben} will show the relationship of the Wald-Iyer formalism
to the present work, where an integral over a surface $\tilde{B}$ rather
than a time slice $\Sigma$ appears. Applying the differential operator
$\left(\square_{{\rm dS}}+d\right)$ to both sides of \ref{eq:mark-wald}
and using the intertwining relation of {[}APPENDIX{]} along with the
equation of motion \ref{eq:hmod-ds-eom} yields the equation \cite{Ben}
\begin{equation}
0=\left(\square_{{\rm dS}}+d\right)\delta S= 2\pi R\left[\delta E_{00}-\delta T_{00}\right].
\end{equation}
Applying $\left(\square_{\mathcal{K}}+2d\right)$ instead yields the
equation
\begin{equation}
0=\left(\square_{\mathcal{K}}+2d\right)\delta S= 2\pi R_{\perp}\left[\delta E_{\mu\nu}-\delta T_{\mu\nu}\right].
\end{equation}
Hence, the equation of motion for $\delta S$ is equivalent to the
linearized gravity equation integrated over a bulk surface, and both
vanish due to the entanglement first law.

The localization of the equation of motion onto the Ryu-Takayanagi
surface after applying these differential operators, a somewhat surprising
fact from the Wald-Iyer point of view, was required by the kinematic
space formalism. It would be interesting to study whether such a localization
occurs more generally away from the vacuum.  If so, it may be more natural to consider whether the bulk gravitational equations can be derived not from entanglement equations, but from one-point entropy equations \cite{Hubeny:2012wa,Kelly:2013aa,Kelly:2014aa}.

%We show that the same result can be achieved with relatively naive (if novel) techniques. 
%Our approach, using intertwinement, establishes an explicit dictionary between the equations of motion for entanglement entropy and the 
%equations of motion for the metric perturbation. 
%In \cite{Ben}, it will be shown that these methods lead to similar equivalence relations for generalized theories of gravity, by building on the Wald-Iyer formalism.

It is also possible to assume \emph{both} the Ryu-Takayanagi formula and Einstein equations, and then derive the kinematic space entropy equations \cite{Nozaki:2013vta,Bhattacharya:2013bna}. One can understand this as a consistency check of the approach, as the entropy equations are pre-determined by conformal invariance.

The techniques we used to derive the quantum corrections for holographic entanglement entropy link this story with a more general program of including interactions in the dynamics of both kinematic space and local bulk operators operators \cite{Kabat:2011rz,Kabat:2015swa,Kabat:2016zzr,Heemskerk:2012mn}. 
In particular, we can think of $\delta A$ as a kinematic field, whose interactions with the stress tensor generate quantum corrections to the kinematic operator.  
We will report on interacting kinematic operators in upcoming work.

\subsection{Acknowledgments}

We thank Jan de Boer, Michal Heller, Rob Myers, Tadashi Takayanagi, and Sandeep Trivedi for useful conversations. 
BC thanks the organizers of the workshop ``Quantum Information in String Theory and Many-body Systems'' at YITP (Kyoto). 
BC, LL, SM thank the Institute for Advanced Study at Tsinghua University (Beijing) for hospitality during the workshop ``An Entangled Trio: Gravity, Information and Condensed Matter.''  BM is supported by The Netherlands Organisation for Scientific Research (NWO).

\appendix

\section{Tensor Radon Transforms}

\subsection{Tensor Intertwinement}\label{Tensor_Int}

The Radon transform maps functions on AdS to functions on the kinematic
space $\mathcal{K}$. In \cite{Czech:2016xec} we derived an \emph{intertwining
relation} for the Radon transform of functions, relating the Laplacian
on AdS-spacetime to the Laplacian on kinematic space: 
\begin{equation}
\square_{\mathcal{K}}Rf=-R\square_{\text{AdS}}f.\label{eq:fintertwine}
\end{equation}
This type of relation has been In this section, we extend the intertwinement
relation (\ref{eq:fintertwine}) to symmetric two-tensors. We will
present a short group theoretic derivation.\\

The only property of the tensor transform that we will use is that
it transforms as a scalar field on kinematic space under the relevant
isometry group. Hence, our proof will hold for a general transform
$\tilde{R}$ with this transformation property. In particular, $\tilde{R}$
can denote the longitudinal transform $R_{\parallel}$, the transverse
transform $R_{\perp}$, or even the wedge transform used in Appendix
\ref{sec:wedge-intertwinement}.

Consider a tensor transform $\tilde{R}\left[s\right]$ of a symmetric
2-tensor field $s_{\mu\nu}$. Under some isometry $g\in{\rm SO}\left(d,2\right)$
of ${\rm AdS}_{d+1}$, this field transforms as $s\rightarrow g\overset{{\rm AdS}}{\cdot}s$,
while a point in kinematic space transforms as $\tilde{B}\rightarrow g\overset{\mathcal{K}}{\cdot}\tilde{B}$.
The scalar transformation property then implies that 
\begin{equation}
\tilde{R}\left[g\overset{{\rm AdS}}{\cdot}s_{\mu\nu}\right]\left(\tilde{B}\right)=\tilde{R}\left[s_{\mu\nu}\right]\left(g^{-1}\overset{\mathcal{K}}{\cdot}\tilde{B}\right).\label{eq:gint}
\end{equation}
In infinitesimal form equation \ref{eq:gint} becomes 
\begin{equation}
L_{AB}^{\left(\mathcal{K},{\rm scalar}\right)}\tilde{R}\left[s_{\mu\nu}\right]=-\tilde{R}\left[L_{AB}^{\left(\text{AdS},\text{2-tensor}\right)}s_{\mu\nu}\right]\label{generator-intertwinement}
\end{equation}
where the $L_{AB}$ are the differential operators representing the
generators of the conformal algebra, and the superscript denotes the
representation. It follows that 
\begin{equation}
L_{\left(\mathcal{K},{\rm scalar}\right)}^{2}\tilde{R}\left[s\right]=\tilde{R}\left[L_{\left(\text{AdS,2-tensor}\right)}^{2}s_{\mu\nu}\right],\label{eq:casimir-tensor-intertwinement}
\end{equation}
where $L^{2}=L_{AB}L^{AB}$ is the conformal Casimir.

To find the intertwining relation for the Laplacian, we must now find
the quadratic differential operator representing $L^{2}$ on the tensor
$s_{\mu\nu}$ and its transform $\tilde{R}\left[s\right]$. To do
this, we make use of the fact that AdS and $\mathcal{K}$ are both
coset spaces $G/H$, where $\text{G}=\text{SO}(d,2)$ and H is the
stabilizer of a bulk point or bulk surface respectively: 
\begin{equation}
\text{AdS}_{d+1}=\frac{\text{SO}\left(d,2\right)}{\text{SO}\left(d,1\right)};\quad\mathcal{K}=\frac{\text{SO}\left(d,2\right)}{\text{SO}\left(d-1,1\right)\times\text{SO}\left(1,1\right)}.
\end{equation}
The Casimir operator on G is represented by the Laplacian
with respect to the Cartan-Killing metric. For a coset space G/H, the Laplacian $\square_{\text{G}}$ can be
written as \cite{pilch1984}: 
\begin{equation}
\square_{\text{G}}=\square_{\text{G/H}}+\square_{\text{H}}.
\end{equation}

A scalar function on AdS-spacetime is in the kernel of the Casimir
of the little group SO$(d,1)$, so the conformal Casimir is represented
on functions on AdS-spacetime by the AdS-Laplacian, up to a constant
proportionality factor. A similar argument holds for functions on
kinematic space. The \emph{relative} proportionality factor in the
intertwining relation \ref{eq:fintertwine} is fixed by a choice of
the Cartan Killing form on the Lie-algebra of the conformal group
$G={\rm SO}\left(d,2\right)$.\footnote{For details on the factor of proportionality, see \cite{Czech:2016xec}.}

For general tensors on AdS-spacetime, there will be an additional
term from the non-trivial representation of the little group H = SO$(d-1,1)$.
The tensor Radon transform maps symmetric (two-) tensors on AdS-spacetime
to \emph{functions} on kinematic space, so there will be no additional
contributions from the Casimir of the kinematic space little group.
Tensors on AdS-spacetime \emph{do} receive a contribution from the
Casimir of the little group H=SO($d,1$). One can decompose a general
tensor on AdS-spacetime in terms of irreducible representations of
the little group H = SO$(d-1,1)$. The irreducible representations
can be labeled by the spin $\ell$, and the \emph{conformal} Casimir
is represented by \cite{pilch1984,Hijano:2015zsa}: 
\begin{equation}
L_{\left(\text{AdS,}\ell\right)}^{2}=-\left(\nabla^{2}+\ell(\ell+d-1)\right).\label{eq:tensor-casimir-representation}
\end{equation}
where $\nabla^{2}$ denotes the covariant Laplacian. We recover the
representation of the conformal Casimir on functions on AdS-spacetime
by setting $\ell=0$. The traceless part of a symmetric two-tensor
corresponds to the $\ell=2$ representation, whereas the trace-part
of a tensor corresponds to the $\ell=0$ representation. We decompose
a general symmetric two-tensor $s_{\mu\nu}$ into the traceless symmetric
and trace parts 
\begin{equation}
s_{\mu\nu}=s_{\mu\nu}^{\text{trace}}+s_{\mu\nu}^{\text{traceless}},\ \ \ s_{\mu\nu}^{{\rm trace}}\equiv\frac{{\rm tr}s}{d+1}g_{\mu\nu}.\label{eq:decomposition}
\end{equation}
Then, using Eqns. \ref{eq:casimir-tensor-intertwinement}, \ref{eq:tensor-casimir-representation},
and \ref{eq:decomposition}, we find the following intertwinement
rule for symmetric two-tensors: 
\begin{align}
\square_{K}\tilde{R}\left[s_{\mu\nu}\right] & =-\tilde{R}\left[\nabla^{2}s_{\mu\nu}^{{\rm trace}}+\left(\nabla^{2}+2\left(d+1\right)\right)s_{\mu\nu}^{{\rm traceless}}\right]\nonumber \\
 & =-\tilde{R}\left[\left(\nabla^{2}+2\left(d+1\right)\right)s-2g_{\mu\nu}{\rm tr}s\right].\label{eq:full-tensor-intertwinement}
\end{align}

\subsection{Einstein Equations from Intertwinement}\label{Einst_Eq}

We would now like to verify Eqn. \ref{eq:einstein-tensor-intertwinement-full}.
First recall from Eqn. \ref{eq:area-perturbation-as-a-transform}
that the area perturbation can be written as 
\begin{equation}
\delta A=\frac{1}{2}R_{\parallel}\left[\delta g_{\mu\nu}\right].
\end{equation}
From here, we can see that the longitudinal Radon transform annihilates
total derivatives:
\begin{equation}
R_{\parallel}\left[\nabla_{\mu}v_{\nu}\right]=0\label{eq:longitudinal-transform-kernel}
\end{equation}
where $v_{\nu}$ is a vector field falling off sufficiently quickly
at infinity that the longitudinal transform is well defined. This
follows from the fact that $\delta A$ vanishes at first order for
small deformations of the surface, which correspond to small coordinate
transformations $\delta g_{\mu\nu}=\nabla_{(\mu}v_{\nu)}$.

Applying the Laplacian $\square_{K}$ to $\delta A$ and using the
intertwining relation \ref{eq:full-tensor-intertwinement}, we find
\begin{equation}
\square_{K}\delta A=-\frac{1}{2}R_{\parallel}\left[\left(\nabla^{2}+2\left(d+1\right)\right)\delta g_{\mu\nu}-2g_{\mu\nu}{\rm tr}\delta g\right]\label{eq:area-intertwinement-step}
\end{equation}
Now, note that the variation of the Ricci tensor is given by 
\begin{multline}
\delta R_{\mu\nu}  =\frac{1}{2}\big[\nabla^{\alpha}\nabla_{\mu}\delta g_{\nu\alpha}+\nabla^{\alpha}\nabla_{\nu}\delta g_{\mu\alpha}\\
  \quad-\nabla^{2}\delta g_{\mu\nu}-\nabla_{\mu}\nabla_{\nu}{\rm tr}\delta g\big]
\end{multline}
By commuting covariant derivatives, this can be rewritten for ${\rm AdS}_{d+1}$
as 
\begin{multline}
\delta R_{\mu\nu} =\frac{1}{2}\bigg[\nabla_{\mu}\left(g^{\alpha\beta}\nabla_{\alpha}\delta g_{\nu\beta}\right)+\nabla_{\nu}\left(g^{\alpha\beta}\nabla_{\alpha}\delta g_{\mu\beta}\right)\\
  -\nabla_{\mu}\nabla_{\nu}{\rm tr}\delta g -\left(\nabla^{2}+2\left(d+1\right)\right)\delta g_{\mu\nu}+2g_{\mu\nu}{\rm tr}\delta g\bigg].
\end{multline}
The last three terms match those in \ref{eq:area-intertwinement-step},
while the longitudinal transform annihilate the first three terms,
yielding the desired result
\begin{equation}
\square_{K}\delta A=R_{\parallel}\delta R_{\mu\nu}.
\end{equation}

Let us now proceed to verify Eqn. \ref{eq:einstein-tensor-intertwinement-slice}.
The Hamiltonian constraint equation, the $tt$ component of the Einstein
equation, can be written as 
\begin{equation}
R^{\prime}-{\rm tr}\left(K^{2}\right)+\left({\rm tr}K\right)^{2}=16 \pi G_N \, T_{00}
\end{equation}
where $R^{\prime}$ and $K_{\mu\nu}$ denote the Ricci scalar and
extrinsic curvature tensors on the time-slice of interest. For an
equal-time slice of AdS, we have $K_{\mu\nu}=0$, so that the linearized
equation takes the simple form
\begin{equation}
\delta R^{\prime}=16 \pi G _N \, \delta T_{00}.
\end{equation}
Hence, we can prove \ref{eq:einstein-tensor-intertwinement-slice}
by showing 
\begin{equation}
\left(\square_{{\rm dS}}+d\right)\delta A=\frac{1}{2}R\left[\delta R^{\prime}\right].\label{eq:slice-wanttoprove}
\end{equation}
If we denote the perturbation of the induced metric on the time-slice
of interest by $\delta w_{\mu\nu}$, then we can write $\delta A=\frac{1}{2}R_{\parallel}\left[\delta w_{\mu\nu}\right]$.
Then, both sides of Eqn. \ref{eq:slice-wanttoprove} depend only on
$\delta w_{\mu\nu}$ , and we can restrict ourselves to considering
perturbations of the metric of hyperbolic space $\mathbb{H}_{d}$.
Generalizing \ref{eq:full-tensor-intertwinement} to hyperbolic space,
we can write the intertwining relation
\begin{equation}
\square_{{\rm dS}}R_{\parallel}\left[\delta w_{\mu\nu}\right]=-R_{\parallel}\left(\left(\nabla^{2}+2d\right)\delta w_{\mu\nu}-2g_{\mu\nu}{\rm tr}\delta w\right).
\end{equation}
Using the same methods as above, we find that 
\begin{equation}
\left(\square_{{\rm dS}}+d\right)\delta A-\frac{1}{2}R\left[\delta R^{\prime}\right]\propto\int_{\tilde{B}}n^{\mu}n^{\nu}\nabla^{\alpha}\nabla_{[\nu}\delta w_{\alpha]\mu}\label{eq:proving-slice-eqn-final-step}
\end{equation}
where $n^{\mu}$
denotes the unit normal vector to $\tilde{B}$ in $\mathbb{H}_{d}$.
To proceed, we must use the fact that the extrinsic curvature tensor
on $\tilde{B}$ is zero, which implies that $\nabla_{\mu}n_{\nu}=n_{\mu}n^{\alpha}\nabla_{\alpha}n_{\nu}$.
Repeatedly using this along with Eqn. \ref{eq:longitudinal-transform-kernel},
we find that the right-hand side of \ref{eq:proving-slice-eqn-final-step}
vanishes, proving the result \ref{eq:einstein-tensor-intertwinement-slice}.

%%%%%%%%%%%%%%%%%%%%%%%%%%%%%%%%%
\section{Wedge Integral Relations\label{sec:wedge-intertwinement}}

In this appendix, we will prove that the equation \ref{eq:correction-equation}
is satisfied by the bulk modular Hamiltonian \ref{eq:bulk-modular-hamiltonian}.
The object of interest will be a transformation $R_{\wedge}\left[s_{\mu\nu}\right]$
of a conserved symmetric 2-tensor $s_{\mu\nu}$, defined by
\begin{equation}
R_{\wedge}\left[s_{\mu\nu}\right]=\int_{\Sigma}s_{\mu\nu}\xi^{\mu}d\Sigma^{\nu}.
\end{equation}
where $\Sigma$ is an equal-time slice of a causal wedge, $d\Sigma^{\nu}$
is the unit normal vector to that slice, and $\xi^{\mu}$ is a Killing
vector given in Poincaré coordinates by \cite{deBoer:2016pqk}
\begin{equation}
\xi^{\mu}=\frac{\left(X-X_{1}\right)^{2}\left(X-X_{2}\right)^{\mu}-\left(X-X_{2}\right)^{2}\left(X-X_{1}\right)^{\mu}}{\left(X_{2}-X_{1}\right)^{2}}\label{eq:killing-vector}
\end{equation}
Here the capital letters denote bulk coordinates $X=\left(x,z\right)$,
whose indices are contracted using the Minkowski metric. The bottom
and top points of the corresponding causal diamond are denoted by
$X_{1}=\left(x_{1},0\right)$ and $X_{2}=\left(x_{2},0\right)$ respectively,
with $\xi$ representing a flow from $X_{2}$ to $X_{1}$. In particular, we have $H_{{\rm bulk}}=2\pi R_{\wedge}\left[T_{\mu\nu}\right]$.
Note that since $s_{\mu\nu}$ is conserved, $R_{\wedge}$ is independent
the choice of time slice $\Sigma$.

We can now restrict ourselves to a constant time slice of ${\rm AdS}_{d+1}$,
which we take to be the $t=0$ slice. Then, $\xi$ points only in
the time direction, and we have
\begin{equation}
R_{\wedge}\left[s_{\mu\nu}\right]=\int_{\Sigma}s_{00}\left|\xi\right|d\Sigma.\label{eq:wedge-transform}
\end{equation}
Hence, this restriction of the wedge transform is really a \emph{scalar}
transform, and is subject to the scalar intertwining relation considered
in \cite{Czech:2016xec}. In particular, we have
\begin{equation}
\left(\square_{{\rm dS}}+d\right)R_{\wedge}\left[s_{\mu\nu}\right]=\int_{\Sigma}\left(-\square_{\mathbb{H}}+d\right)s_{00}\left|\xi\right|.
\end{equation}
Integrating by parts, this becomes 
\begin{multline}
\left(\square_{{\rm dS}}+d\right)R_{\wedge}\left[s_{\mu\nu}\right] =-\int_{\partial\Sigma}\left|\xi\right|n^{\mu}\nabla_{\mu}s_{00}+\\
\int_{\Sigma}s_{00}\left(d-\nabla^{2}\right)\left|\xi\right| +\int_{\partial\Sigma}s_{00}n^{\mu}\nabla_{\mu}\left|\xi\right|
\end{multline}
where $n^{\mu}$ is the outward-pointing unit normal vector to $\tilde{B}$
within the specified time slice. The first term vanishes because $\left|\xi\right|=0$
at $\tilde{B}$, and because $s_{00}$ goes to zero sufficiently quickly
at $\partial{\rm AdS}$; the second vanishes since $\nabla_{\mathbb{H}}^{2}\left|\xi\right|=d\left|\xi\right|$,
as can be checked explicitly from \ref{eq:killing-vector}. Finally,
we can check from \ref{eq:killing-vector} that $n^{\mu}\nabla_{\mu}\left|\xi\right|=-1$.
This gives us our result,
\begin{equation}
\left(\square_{{\rm dS}}+d\right)R_{\wedge}\left[s_{\mu\nu}\right]=-R\left[s_{\mu\nu}\hat{t}^{\mu}\hat{t}^{\nu}\right]
\end{equation}
 where $\hat{t}$ is the timelike unit vector to the time slice correponding
to the chosen de Sitter slice of kinematic space.

With some more effort, we can also prove a similar relation using
the Laplacian on the full kinematic space, rather than a de Sitter
slice. First, note that the wedge integral can be written as 
\begin{equation}
R_{\wedge}\left[s_{\mu\nu}\right]=\int_{\Sigma}\ast j.
\end{equation}
where we have defined the conserved current
\begin{equation}
j_{\mu}=s_{\mu\nu}\xi^{\nu}.
\end{equation}
Now, since $R_{\wedge}$ transforms as a scalar in the full kinematic
space, we can use the tensor intertwining relation \ref{eq:full-tensor-intertwinement}
to obtain
\begin{align}
\left(\square_{K}+2d\right)R_{\wedge}\left[s_{\mu\nu}\right] & =R_{\wedge}\left[-\left(\nabla^{2}+2\right)s_{\mu\nu}+2g_{\mu\nu}{\rm tr}s\right]\nonumber\\
 & =\int_{\Sigma}\ast\tilde{j}
\end{align}
where $\tilde{j}_{\mu}=\left[-\left(\nabla^{2}+2\right)s_{\mu\nu}+2g_{\mu\nu}{\rm tr}s\right]\xi^{\nu}$.
Using the fact that $s_{\mu\nu}$ is conserved and $\xi^{\mu}$ is
Killing, it can be shown with significant effort \cite{Ben} that $\tilde{j}=\Delta j$,
where $\Delta$ is the Hodge Laplacian. Then, conservation of $j$
implies that 
\begin{align}
\left(\square_{K}+2d\right)R_{\wedge}\left[s_{\mu\nu}\right] & =-\int_{\tilde{B}}\ast dj.\nonumber\\
 & =-\frac{1}{2}\int_{\tilde{B}}\epsilon^{\mu\nu}\left(dj\right)_{\mu\nu}
\end{align}
where $\epsilon^{\mu\nu}$ is the antisymmetric tensor in the two
directions perpendicular to $\partial\Sigma$, defined such that $\epsilon^{01}=-1$.
Next, note that since $\xi$ vanishes on $\tilde{B}$, we can plug
in $j$ to find 
\begin{equation}
\left(\square_{K}+2d\right)R_{\wedge}\left[s_{\mu\nu}\right]=\int_{\tilde{B}}\left(\epsilon^{\mu\alpha}\nabla_{\alpha}\xi^{\nu}\right)s_{\mu\nu}.
\end{equation}
Finally, it can be checked explicitly from \ref{eq:killing-vector}
that $\epsilon^{\mu\alpha}\nabla_{\alpha}\xi^{\nu}=g^{\mu\nu}-h^{\mu\nu}$,
where $h^{\mu\nu}$ is the induced metric on $\tilde{B}$. This yields
the result
\begin{equation}
\left(\square_{K}+2d\right)R_{\wedge}\left[s_{\mu\nu}\right]=R_{\perp}\left[s_{\mu\nu}\right].
\end{equation}
This relation was required for consistency between equations \ref{eq:area-perturbation-ks-eqn},
\ref{eq:hmod-full-eom}, and \ref{eq:full-kinematic-dictionary}.

\section{Modular Hamiltonian as an OPE Block}\label{ModHam}

In this appendix, we will relate the stress tensor OPE block to the
vacuum modular Hamiltonian for a spherical CFT region. This implies
that the vacuum modular Hamiltonian appears as the contribution of
the stress tensor to the OPE of timelike separated scalars of equal
dimension, or the expansion of a spherical operator as in Eq. \ref{eq:sphere-operator-blocks}.

In the OPE of two timelike separated scalars $O\left(x\right)$ of
equal scaling dimension $\Delta$, the stress tensor and its derivatives
appear as 
\begin{equation}
\frac{O\left(x_{1}\right)O\left(x_{2}\right)}{\left\langle O\left(x_{1}\right)O\left(x_{2}\right)\right\rangle }\supset C_{OOT}\underbrace{{\scriptstyle \left(1+\ldots\right)\frac{x_{12}^{\mu}x_{12}^{\nu}}{x_{12}^{2-d}}}T_{\mu\nu}\left({\scriptstyle \frac{x_{1}+x_{2}}{2}}\right)}_{{\textstyle \mathcal{B}_{T}\left(x_{1},x_{2}\right)}}.\label{eq:tensor-ope-block}
\end{equation}
Here, we have defined the OPE block $\mathcal{B}_{T}$ for the stress
tensor, which includes the contribution of $T_{\mu\nu}$ and all descendants
to the OPE, and is independent of the choice of operator $O\left(x\right)$.
Note that, when expanded about the point  $\frac{x_{1}+x_{2}}{2}$,
we can use tracelessness and conservation to ensure that only the
quantity $x_{12}^{\mu}x_{12}^{\nu}T_{\mu\nu}$ and its derivatives
in directions perpendicular to $x_{12}$ appear. More concretely,
if we choose $x_{2}=-x_{1}=R\hat{t}$, this means that only $T_{00}$
and its spatial derivatives appear, as we would expect from the expression
\ref{eq:Hmod}.

Now, consider the ${\rm SO}\left(d,1\right)$ subgroup of the conformal
group which preserves the time slice intersecting $\frac{x_{1}+x_{2}}{2}$
and the sphere corresponding to $x_{1},x_{2}$. The quantity $x_{12}^{\mu}x_{12}^{\nu}T_{\mu\nu}$
transforms as a scalar primary of dimension $d$ under this subgroup,
so it has eingenvalue $-\Delta\left(\Delta-\left(d-1\right)\right)=-d$
under the Casimir operator $L_{{\rm SO}\left(d,1\right)}^{2}$, as
do its derivatives in directions along the time slice. Since $\mathcal{B}_{T}\left(x_{1},x_{2}\right)$
transforms as a scalar in kinematic space, $L_{{\rm SO}\left(d,1\right)}^{2}$
is represented by $\square_{{\rm dS}}$, the Laplacian on the de
Sitter slice of kinematic space corresponding to this ${\rm SO}\left(d,1\right)$
subgroup \cite{PhysRevLett.116.061602}. Hence, $\mathcal{B}_{T}$ obeys the equation of motion
\begin{equation}
\left(\square_{{\rm dS}}+d\right)\mathcal{B}_{T}=0,
\end{equation}
which of course matches Eqn. \ref{eq:hmod-ds-eom} for the modular
Hamiltonian; in fact, we obtain a whole family of such equations,
one for each time slice. Comparing boundary conditions of \ref{eq:hmod-boundary-conditions}
and \ref{eq:tensor-ope-block}, we obtain the relation 
\begin{equation}
\mathcal{B}_{T}=-\frac{2^{d}\left(d^{2}-1\right)}{2\pi\Omega_{d-2}}H_{{\rm mod}}
\end{equation}
where $\Omega_{d-2}$ is the area of a $\left(d-2\right)$-sphere.
This result can also be obtained through the shadow operator formalism
\cite{Czech:2016xec}.\\

\bibliographystyle{jhep}
\bibliography{bibliography}

\end{document}